\documentclass[aps,prb,twocolumn]{revtex4}
\usepackage{graphicx}
\usepackage{dcolumn}
\usepackage{amsmath}
\usepackage{amssymb}
\usepackage{float}
\usepackage{epstopdf}
\def\ud{\uparrow\downarrow}

\def\d{\downarrow}
\def\u{\uparrow}
\def\eclr{\epsilon_c^{\rm LR}}
\def\erfc{{\rm erfc}}
\def\erf{{\rm erf}}
\def\rv{{\bf r}}

\def\xv{{\bf x}}

\def\ec{\epsilon_c}

\def\beq{\begin{equation}}
\def\eeq{\end{equation}}

\def\kf{k_{\rm F}}

\begin{document}
\title{Local-spin-density functional for multideterminant density functional theory}
\author{Simone Paziani,$^1$ Saverio Moroni,$^2$ Paola Gori-Giorgi,$^3$ 
and Giovanni B. Bachelet$^1$}

\affiliation{
$^1$INFM Center for Statistical Mechanics and Complexity and Dipartimento di Fisica, Universit\`a di Roma ``La Sapienza'', Piazzale A. Moro 2, I-00185 Roma, Italy \\
$^2$INFM DEMOCRITOS National Simulation Center,
    via Beirut 2--4, I--34014 Trieste, Italy\\
$^3$Laboratoire de Chimie Th\'eorique, CNRS UMR7616,
    Universit\'e Pierre et Marie Curie, 4 Place Jussieu,
    F-75252 Paris, France
}
\date{\today}
\begin{abstract}
Based on exact limits and quantum Monte Carlo simulations, we obtain, at any density and spin polarization, an accurate estimate for the energy of a modified homogeneous electron gas where electrons repel each other only with a long-range coulombic tail.
This allows us to construct an analytic local-spin-density exchange-correlation functional appropriate to new, multideterminantal versions of the density functional theory, where quantum chemistry and approximate exchange-correlation functionals are combined to optimally describe both long- and short-range electron correlations.
\end{abstract}

\maketitle
\section{Introduction}
Density functional theory \cite{kohnnobel,science,FNM} (DFT) 
is by now the most popular method for electronic structure calculations
in condensed matter physics and quantum chemistry, because of its unique combination of low computational cost and high accuracy for many molecules and solids. There are, however, exceptions to such an accuracy. 
Even the best approximations of its key ingredient, the exchange-correlation (XC) energy functional, cannot describe strong electron correlations, like those of the cuprates, and cannot exactly cancel the so-called self-interaction, a property which the exact functional should satisfy. On top of that, they fail  to recover long-range van der Waals interactions,\cite{notavdW} are not completely safe for the description of the hydrogen bond,\cite{fuchs2} and have intrinsic problems with situations of
near degeneracy (when two sets of Kohn-Sham orbitals have very close energies);\cite{erf,erf1,conLDA} more generally, the ``chemical accuracy'' has not yet been reached. To overcome the latter group of problems, there has been a growing interest in ``mixed schemes'' which combine the DFT with other approximate methods by splitting the coulombic electron-electron interaction $1/r=v_{ee}(r)$ into a short-range (SR) and a long-range (LR) part (see e.g. Refs.~\onlinecite{kohn_mm,ikura,scuseria,vaffa,erf,erf1,conLDA,julGGA,sav_madrid,julien,janos,
stoll,yamaguci}). 
The idea is to use different approximations for the LR and
the SR contributions to the exchange and/or correlation energy
density functionals of the Kohn-Sham (KS) scheme. It descends from
the observation that LR correlations, poorly described by local or semi-local density functionals, can be accurately dealt with
by other techniques, like the random-phase approximation (RPA) or standard wavefunction methods of quantum chemistry. 
Conversely, correlation effects due to the SR part of the 
electron-electron interaction are in general 
well described by local or semilocal functionals.\cite{BPE,ontop} The error function and its complement 
\beq
\frac{1}{r}=v_{ee}(r)=v_{\rm SR}^\mu(r)+v_{\rm LR}^\mu(r)=
\frac{\erfc(\mu r)}
{r}+\frac{\erf(\mu r)}{r}
\label{eq_split}
\eeq
where $\mu$ controls the range of the decomposition, are often used\cite{ikura,erf,erf1,conLDA,julGGA,scuseria,julien,janos,stoll} 
to split the Coulomb interaction into a SR and a LR part, 
since they yield analytic matrix elements for both Gaussians 
and plane waves, i.e., the most common basis functions in quantum chemistry 
and solid state physics. Correspondingly, the universal functional of
the electron density $n$, as defined in the constrained-search formalism,\cite{mel}
\beq
F[n]=\min_{\Psi\to n}\langle \Psi|T+V_{ee}|\Psi\rangle.
\label{eq_stDFT}
\eeq
can be divided into a short-range and a complementary long-range part $F[n]  =  F_{\rm SR}^\mu[n]+\overline{F}_{\rm LR}^\mu[n]$
\begin{eqnarray}
F_{\rm SR}^\mu[n] & = & \min_{\tilde{\Psi}^\mu\to n}\langle 
\tilde{\Psi}^\mu|T+V_{\rm SR}^\mu|\tilde{\Psi}^\mu\rangle 
\nonumber \\
\overline{F}_{\rm LR}^\mu[n] & = & F[n]-F_{\rm SR}^\mu[n].
\label{eq_SR-LR}
\end{eqnarray}
or, alternatively, into a long-range and a complementary
short-range part $F[n]  =  F_{\rm LR}^\mu[n]+\overline{F}_{\rm SR}^\mu[n]$
\begin{eqnarray}
F_{\rm LR}^\mu[n] & = & \min_{\Psi^\mu\to n}\langle \Psi^\mu|T+V_{\rm LR}^\mu|\Psi^\mu\rangle 
\nonumber \\
\overline{F}_{\rm SR}^\mu[n] & = & F[n]-F_{\rm LR}^\mu[n].
\label{eq_LR-SR}
\end{eqnarray}
The two decompositions lead to different strategies and XC energy functionals, whose merits and drawbacks are discussed in Ref.~\onlinecite{jul-sav}. In any event, for actual electronic-structure calculations to be performed, these functionals ultimately need approximations, in analogy with the standard DFT. Regardless of the strategy adopted, the potential superiority of  ``mixed schemes" comes into play precisely at this stage: compared to the standard version, a DFT which only deals with the SR part of the electron-electron interaction should be much more accurately approximated, as mentioned, by local-density XC energy functionals.\cite{BPE,ontop,julien}
While both decompositions [Eq.~(\ref{eq_SR-LR}) and Eq.~(\ref{eq_LR-SR})] are aimed at the exploitation of the DFT scheme for the SR part of the interaction only, the corresponding approximate functionals require an accurate description of the homogeneous electron gas (HEG) either with SR [Eq.~(\ref{eq_SR-LR})] or LR interaction [Eq.~(\ref{eq_LR-SR})].
\par
Up to now, the HEG exchange-correlation energies as a function of the cutoff
parameter $\mu$ and of the electron density are available for the SR case 
from  Quantum Monte Carlo (QMC) simulations,\cite{Zecca_04} 
and for the LR case from coupled-cluster (CC) calculations.\cite{erf1,erfgau}  
A parametrization of the CC data for the XC energy of the HEG with long-range-only interaction has been used 
in Refs.~\onlinecite{conLDA} and~\onlinecite{janos} with very 
promising results for
closed-shell systems. Generalized-gradient-corrected density
functionals have also been designed and tested within this 
framework,\cite{julGGA,stoll} but all existing functionals are limited to the spin-unpolarized case.
\par
The purpose of this paper is to provide, based on novel exact limits and quantum Monte Carlo simulations, an accurate representation for the energy of the LR-only interacting HEG not only as a function of the cutoff  parameter $\mu$ and of the total electron density, but also as a
function of the spin polarization [i.e., as a function of the spin densities
$n_\uparrow(\rv)$ and $n_\downarrow(\rv)$ separately]. 
Since von Barth and Hedin\cite{Ulf} showed, in 1972, that the task of finding good approximations to exchange-correlation density functionals is greatly simplified if the functional is expressed in terms of the spin densities, and that this is the simplest way to satisfy the requirement (Hund's rule) that a state with larger spin tends to be energetically favored, the importance of including such a spin dependence in approximate functionals was confirmed by countless calculations for molecules and solids.\cite{Olle, nagyreview} 
In this context the decomposition of Eq.~(\ref{eq_LR-SR}), based on the constrained-search formalism,\cite{mel} is generalized to spin-DFT as follows:\cite{nagyreview} 
\begin{eqnarray}
F_{\rm LR}^\mu[n_{\uparrow},n_{\downarrow}] & = & \min_{\Psi^\mu\to n_{\uparrow},n_{\downarrow}}\langle \Psi^\mu|T+V_{\rm LR}^\mu|\Psi^\mu\rangle 
\nonumber \\
\overline{F}_{\rm SR}^\mu[n_{\uparrow},n_{\downarrow}] & = & F[n_{\uparrow},n_{\downarrow}]-F_{\rm LR}^\mu[n_{\uparrow},n_{\downarrow}].
\label{eq_LR-SR-spin}
\end{eqnarray}
The spin-polarized LR-only gas, for which no previous results are to our knowledge available, is appropriate, via Eq.~(\ref{eq_LR-SR-spin}), 
to a very promising ``multideterminantal'' version of the spin-DFT.
The final outcome of this work is thus a local-spin-density approximation of the corresponding XC functional, given in analytic form, by which 
electronic structure calculations of this new type will be possible for unpolarized systems\cite{notaCC} and, more important,
for spin-polarized systems, for which no such functional is presently available. 
Such a functional also represents the key ingredient for extending gradient-corrected SR density functionals\cite{julGGA,stoll} to spin-DFT.
 \par
The paper is organized as follows. In Sec.~\ref{sec_def} we define the
hamiltonian of the HEG with LR-only interaction and we derive some
exact limits of the corresponding correlation energy, which is then
computed (for values of the relevant parameters not accessible 
to analytic methods) with QMC in Sec.~\ref{sec_dmc}. The
results of Secs.~\ref{sec_def} and \ref{sec_dmc} are then used in
Sec.~\ref{sec_fit} to construct an analytic parametrization of the
LR correlation energy. Sec.~\ref{sec_delta} recalls, calculates and provides in analytic form an alternative definition of the LR correlation energy which involves the use of pair-correlation functions (also obtained from our QMC simulations) and may be of interest within optimized-effective-potential schemes.\cite{TGS2} \par Hartree atomic units are used throughout this work.

\section{Definitions, and exact limits}
\label{sec_def}
After decomposing the standard (Coulomb interaction) spin-DFT functional 
according to Eq.~(\ref{eq_LR-SR-spin}),  the resulting SR functional $\overline{F}_{\rm SR}^\mu[n_\uparrow,n_\downarrow]$ can be further decomposed, as usually, into a Hartree and an XC term
\beq
\overline{E}_{\rm H}^\mu[n]= 
{E}_{\rm H}^\mu[n]=
\frac{1}{2}\int \!\! d\rv \!\! \int \!\!
d\rv' n(\rv)\,n(\rv')\,v^\mu_{\rm SR}(|\rv-\rv'|)
\label{eq_EHcomp}
\eeq
\beq
\overline{E}_{\rm xc}^\mu[n_\uparrow,n_\downarrow]=
\overline{F}_{\rm SR}^\mu[n_\uparrow,n_\downarrow]
-\overline{E}_{\rm H}^\mu[n].
\label{XCdefinit}
\eeq
The local-spin-density (LSD) approximation amounts to replacing the exact, unknown functional of Eq.~(\ref{XCdefinit}) with
\begin{eqnarray}
&&\overline{E}_{\rm xc,LSD}^\mu[n_\uparrow,n_\downarrow]= 
\label{locspindens} \\
&&\int d\rv\,n(\rv)\, [\epsilon_{\rm xc}(n_\uparrow(\rv),n_\downarrow(\rv))-
\epsilon_{\rm xc}^{\rm LR}(n_\uparrow(\rv),n_\downarrow(\rv),\mu)]=\nonumber\\
&& \int d\rv\,n(\rv)\, [\epsilon_{\rm xc}(r_s(\rv),\zeta(\rv))-
\epsilon_{\rm xc}^{\rm LR}(r_s(\rv),\zeta(\rv),\mu)],
\nonumber
\end{eqnarray}
where $\epsilon_{\rm xc}(n_\uparrow,n_\downarrow)$ is the exchange-correlation energy per particle of the standard jellium model\cite{CA,VWN,PZ,PW92} with uniform spin densities $n_\uparrow,n_\downarrow$, and
$\epsilon_{\rm xc}^{\rm LR}(n_\uparrow,n_\downarrow,\mu)$ is the
corresponding quantity for a jellium model with LR-only interaction $v_{\rm LR}^\mu(r)$, which forms the object of this paper. In the third line of
Eq.~(\ref{locspindens}) we express the same quantity in terms of the electronic density $n=n_\uparrow\!+\!n_\downarrow=(4\pi r_s^3/3)^{-1}$ and spin polarization $\zeta=(n_\u-n_\d)/n$, thus introducing the notation used in what follows.
To obtain $\epsilon_{\rm xc}^{\rm LR}(r_s,\zeta, \mu)$ we consider a uniform system with LR-only interaction
\begin{equation}
H^{\mu}_{\rm LR}  =  -\frac{1}{2}\sum_{i=1}^N \nabla^2_{\rv_i}+V_{\rm LR}^\mu
+V_{eb}^{\mu}+V_{bb}^{\mu},
\label{eq_ham}
\end{equation}
where $V_{\rm LR}^\mu$ is the modified electron-electron interaction
\begin{equation}
V_{\rm LR}^\mu  =  \frac{1}{2}\sum_{i\ne j=1}^N\frac{\erf(\mu|\rv_i-\rv_j|)}
{|\rv_i-\rv_j|}, 
\label{eq_Vee}
\end{equation}
$V_{eb}^{\mu}$ is the interaction between the electrons and a rigid,
positive, uniform background of density $n$
\begin{equation}
V_{eb}^{\mu}  =  -n\sum_{i=1}^N\int d\xv \,\frac{\erf(\mu|\rv_i-\xv|)}
{|\rv_i-\xv|}, 
\label{eq_Veb}
\end{equation}
and $V_{bb}^{\mu}$ is the corresponding background-background interaction
\begin{equation}
V_{bb}^{\mu}  =  
\frac{n^2}{2}\int d\xv\int d\xv' \,\frac{\erf(\mu|\xv-\xv'|)}
{|\xv-\xv'|}.
\label{eq_Vbb}
\end{equation}
Our hamiltonian $H^{\mu}_{\rm LR}$,
and thus its ground-state energy per electron
$\epsilon^{\rm LR}$, depends on the density
parameter $r_s$, on the spin-polarization $\zeta$,
and on the cutoff parameter $\mu$.
When $\mu\to\infty$, we recover the standard jellium model; in the opposite limit $\mu\to 0$, we recover the
noninteracting electron gas. 
In this Section we derive the asymptotic behavior for $\mu\to 0$ and $\mu\to\infty$ of the correlation energy
per electron, defined as \begin{eqnarray}
&\eclr(r_s,\zeta,\mu)&=
\label{eq_standardec}\\
&&\epsilon^{\rm LR}(r_s,\zeta,\mu)-t_s(r_s,\zeta)-
\epsilon_x^{\rm LR}(r_s,\zeta,\mu),
\nonumber 
\end{eqnarray}
where $t_s(r_s,\zeta)=3\kf^2\,\phi_5(\zeta)/10$ 
is the usual kinetic energy of the noninteracting
electron gas, with $\kf=(\alpha\, r_s)^{-1}$,
$\alpha=(4/9\pi)^{1/3}$, and 
\beq
\phi_n(\zeta)=\frac{1}{2}\left[ (1\!+\!\zeta)^{n/3}+(1\!-\!\zeta)^{n/3} \right];
\label{eq_phi}
\eeq
the
exchange energy is given by\cite{erf1}
\begin{eqnarray}
\epsilon_x^{\rm LR}(r_s,\zeta,\mu) & = & \frac{1}{2}(1\!+\!\zeta)^{4/3}
f_x\left(r_s,\mu(1\!+\!\zeta)^{-1/3}\right)+ \nonumber \\
& + & \frac{1}{2}(1\!-\!\zeta)^{4/3}
f_x\left(r_s,\mu(1\!-\!\zeta)^{-1/3}\right),\label{eq_ex} \\
f_x(r_s,\mu) & = & -\frac{\mu}{\pi}\biggl[(2y-4y^3)\,e^{-1/4y^2}-
3y+4y^3+ \nonumber \\
& + & \sqrt{\pi}\,\erf\left(\frac{1}{2y}\right)\biggr],
\qquad y=\frac{\mu\,\alpha\,r_s}{2},
\end{eqnarray}
and has the asymptotic behaviors
\begin{eqnarray}
\epsilon_x^{\rm LR}(r_s,\zeta,\mu)\Big|_{\mu\to 0}  =
-\frac{\mu}{\sqrt{\pi}}+\frac{3\alpha r_s \mu^2}{2\pi}\,\phi_2(\zeta)+
O(\mu^3) \label{eq_exsmallmu}
\\
\epsilon_x^{\rm LR}(r_s,\zeta,\mu)\Big|_{\mu\to \infty}  \!\!\!\!\!\!\!\!\!\! =
-\frac{3\kf}{4\pi}\phi_4(\zeta) 
+ \frac{3\,(1\!+\!\zeta^2)}{16\,r_s^3\,\mu^2}+O(\mu^{-4})
\end{eqnarray}

\subsection{Approaching the non-interacting gas}
When $\mu\to 0$ and/or $r_s\to 0$, we are approaching the limit of
the non-interacting Fermi gas. Toulouse {\it et al.}\cite{julien} 
have studied the $\mu\to 0$ limit of the long-range exchange and correlation 
energy functionals for confined systems (atoms, molecules) using standard perturbation theory. 
Their results cannot be applied to the case of an extended system like 
the uniform electron gas, because the integrals of their Eqs.~(17) and (20)
would diverge.
Instead, the $\mu\to 0$ limit (as well as the $r_s\to 0$ limit) 
of the uniform electron gas can be studied with RPA,\cite{pines,Ulf} which becomes exact both for long-range correlations ($\mu\to 0$: in this limit the
  long-range coulombic tail shows up only beyond larger and larger
  interelectronic distance $r \sim 1/\mu$) and in the high-density
  limit ($r_s\to 0$). We generalize to the LR-only interaction $\erf(\mu r)/r$ the standard RPA expression for the correlation energy 
(see Appendix~\ref{app_RPA} for details), and find that, 
for small $\mu\sqrt{r_s}$ (i.e., small-$\mu$ and/or
      $r_s\to 0$ limit), the correlation energy $\eclr$ scales as
\beq
 \eclr(r_s,\zeta,\mu)\Big|_{\mu,r_s\to 0}\!\!\!\!\!= \left[\phi_2(\zeta)\right]^3\,Q(x), \quad
 x=\frac{\mu \sqrt{r_s}}{\phi_2(\zeta)},  
\label{eq_smallmu}
\eeq
where $\phi_2(\zeta)$ is given by Eq.~(\ref{eq_phi}), and
the function $Q(x)$ has the following asymptotic behaviors
\begin{eqnarray}
Q(x\to 0) & = & -\frac{3\alpha}{2\pi}x^2+O(x^3) \label{eq_smallx} \\
Q(x\to\infty) & = & \frac{2\ln(2)-2}{\pi^2}\ln(x)+{\rm const.}
\label{eq_largex}
\end{eqnarray}
The scaling of Eq.~(\ref{eq_smallmu}) for the long-range
correlation energy was also expected from the fact that the long-range
part of the pair-correlation function of the standard jellium model
has a similar scaling.\cite{WP91,PW,GP2}
Notice also that, in the small-$\mu$ expansion of
$\eclr$ [Eqs.~(\ref{eq_smallmu}) and (\ref{eq_smallx})],
the term proportional to $\mu^2$ exactly cancels with the corresponding
term in the exchange energy $\epsilon_x^{\rm LR}$ [Eq.~(\ref{eq_exsmallmu})], so
that the XC energy (exchange {\em plus} correlation) has no $\mu^2$ term; in a confined system, on the other hand, the $\mu^2$ terms are separately zero for exchange and correlation.\cite{julien}
We found that the function $Q(x)$ (see Appendix~\ref{app_RPA})
is accurately approximated by
\beq
Q(x)=\frac{2\ln(2)-2}{\pi^2}\ln\left(\frac{1+a\,x+b\,x^2+c\,x^3}{1+a\,x+
d\,x^2}\right),
\label{eq_Q}
\eeq
with $a=5.84605$, $c=3.91744$, $d=3.44851$, and $b=d-3\pi\alpha/(4\ln(2)-4)$.
\par
A final remark on the scaling of Eq.~(\ref{eq_smallmu}) is that,
although it exactly holds only in the small-$x$ regime, even in the
large-$x$ regime of Eq.~(\ref{eq_largex}) (obtained e.g. with small
$r_s$ but  very large $\mu$), it represents an excellent approximation,
because in this regime the $\zeta$ dependence of $\eclr/Q$ is described
by a function [Eq.~(32) of Ref.~\onlinecite{WPHDsc}] which is very
similar to $[\phi_2(\zeta)]^3$, and exactly equals it for
$\zeta\!=\!0$ and $\zeta\!=\!1$.
All the densities corresponding to $r_s\gtrsim 0.1$ are not affected 
by this small difference in the $\zeta$ dependence of
Eq.~(\ref{eq_smallmu}), as discussed in Appendix~\ref{app_RPA}.

\subsection{Approaching the coulombic gas}
The large-$\mu$ behavior of the long-range correlation functional
appropriate to the decomposition of Eq.~(\ref{eq_LR-SR-spin}), obtained
in Refs.~\onlinecite{julien} and \onlinecite{GS3}, is straigthforwardly
extended to the uniform electron gas. For $\zeta\neq 1$ we
have\cite{GS3}
\begin{eqnarray}
\eclr(r_s,\zeta,\mu)\Big|_{\mu\to\infty} & = & \ec(r_s,\zeta)
-\frac{3\,g_c(0,r_s,\zeta)}{8\,r_s^3\,\mu^2} \nonumber \\
& - & \frac{g(0,r_s,\zeta)}{\sqrt{2\pi}\, r_s^3\,\mu^3}
+O(\mu^{-4})
\label{eq_mugraz0}
\end{eqnarray}
where $\ec(r_s,\zeta)$ is the correlation energy of the
standard electron gas with Coulomb interaction, 
$g(0,r_s,\zeta)$ its on-top pair-distribution function,~\cite{PW,GP2,GP1}
and $g_c(0,r_s,\zeta)=g(0,r_s,\zeta)-\frac{1}{2}(1-\zeta^2)$.
For the fully polarized gas ($\zeta\!=\!1$) the terms proportional  to $\mu^{-2}$ and $\mu^{-3}$ in the large-$\mu$ expansion 
of $\eclr$ vanish, and the next leading terms
are\cite{GS3}
\begin{eqnarray}
\eclr(r_s,\zeta\!=\!1,\mu)\Big|_{\mu\to\infty} & = & \ec(r_s,\zeta\!=\!1)
-\frac{9\,g_c''(0,r_s,\zeta\!=\!1)}{64\,r_s^3\,\mu^4} \nonumber \\
& - & \frac{9\, g''(0,r_s,\zeta\!=\!1)}{40\sqrt{2\pi}\,r_s^3\,\mu^5}
+O(\mu^{-6})
\label{eq_mugraz1}
\end{eqnarray}
where $g''(0,r_s,\zeta\!=\!1)$ is the second derivative at $r=0$ of
the pair-distribution function\cite{PW,GP1,GP2} of the fully polarized
gas, and $g_c''(0,r_s,\zeta\!=\!1)=g''(0,r_s,\zeta\!=\!1)-2^{5/3}\kf^2/5$.

\section{Diffusion Monte Carlo}
\label{sec_dmc}
The details of the simulations are similar to our previous calculation
of a local density functional for a short-range potential.\cite{Zecca_04}
Here we give a technical summary focusing on the main differences, 
which concern the size extrapolation and the treatment of the 
long-range tails of the interaction and of the pair pseudopotential 
in the trial wave function. For the reader not keen on
technicalities, it is enough to say that we provide a very tight upper 
bound to the exact ground-state energy, choosing a level of 
approximation which closely
matches the Ceperley--Alder\cite{CA} (CA) result for $\mu\to\infty$. 

The ground-state energy of the Hamiltonian of Eq.~(\ref{eq_ham})
is computed with the diffusion Monte Carlo (DMC) method in the fixed-node
(FN) approximation,\cite{Foulkes_01} using a standard Jastrow--Slater
trial function with plane--wave orbitals and RPA 
pseudopotentials.\cite{Ceperley_78}
Several values of the density ($r_s\!=\!1,2,5,$ and 10), of the cutoff
parameter ($\mu r_s=0.5, 1, 2$ and 4) and of the spin polarization
($\zeta\!=\!0$ and $1$) are considered. 
The results are fitted (see Sec.~\ref{sec_fit}) to a 
convenient analytical expression for the correlation energy 
$\epsilon_c(r_s,\mu,\zeta)$, which also embodies the exact
limits of Sec.~\ref{sec_def} and is further constrained to recover the 
CA result for the Coulomb potential.

This constraint sets the target precision of our simulations, since 
there is no point in pushing the accuracy much beyond
the statistical uncertainty of the CA results. Correspondingly, we make
sure that the biases due to a finite time step and a finite number of walkers 
are much smaller than the statistical uncertainties of the CA results.
Furthermore, as discussed in Ref.~\onlinecite{Zecca_04},
a smoother match to the CA results of the FN energy
in the $\mu\to\infty$ limit is expected using the nodal structure given by 
Slater determinants of plane waves, instead of the more 
accurate\cite{Kwon_98} (and computationally more demanding) 
backflow nodes.

We simulate $N$ particles in a cubic box with ``twist--averaged boundary
conditions''\cite{Lin_01} (TABC), which have been shown to eliminate most
of the finite--size effect due to the shell structure of the plane--wave
determinants. For each system considered, simulations are performed for
35 points in the irreducible wedge of the first Brillouin zone (BZ) of the
simulation box, corresponding to a 1000--point mesh in the whole BZ.

Both the interparticle potential and the RPA pair pseudopotential are
computed using an optimized breakup\cite{Natoli_95}
into a long--range part, to be treated in reciprocal space, and 
a short--range part, to be treated in real space.
The short--range part is expanded in locally piecewise quintic Hermite
interpolants over 20 knots, and the $k$--space summation includes
20 shells of reciprocal lattice vectors.
This choice of parameters ensures that, for the Coulomb interaction, 
the potential energy calculated for
a simulation box containing 64 particles on a simple cubic lattice 
reproduces the exact Madelung constant to less than 1 part in 10$^7$.
\begin{figure}
\includegraphics[width=5.6cm,angle=270]{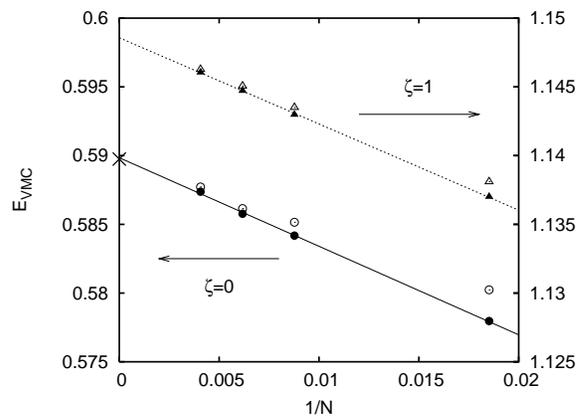} 
\caption{The size dependence of the VMC energy for the 
Coulomb potential at $r_s\!=\!1$. Empty symbols: VMC energy $E_N$; 
filled symbols: $E_N + T_\infty - T_N$. The curves
show the best--fit $1/N$ dependence of the latter, according to
Eq.~\ref{eq:size_extrapolation}:
circles and solid line refer to $\zeta\!=\!0$ (left scale), triangles 
and dashed line to $\zeta\!=\!1$ (right scale). The cross is the $\zeta\!=\!0$ result obtained in the thermodynamic limit by Ref.~\onlinecite{Kwon_98}, which
appears fully consistent with the present calculation. The
statistical errors on the data points are much smaller than the
symbol sizes. The $\chi^2$ is 11.8 for $\zeta\!=\!0$, and 0.4 for $\zeta\!=\!1$ (much
poorer values would be obtained by just fitting $E_{VMC}$, i.e., without including the kinetic-energy size correction).}
\label{fig:size_extrapolation}
\end{figure}

All DMC simulations have been
done with $N=54$ for both the paramagnetic and the spin--polarized fluids
(there is no need of choosing closed--shell determinants with TABC). 
Following a common practice,\cite{CA,Kwon_98} the residual size 
effect has been estimated assuming that it is the same 
for DMC and Variational Monte Carlo (VMC),\cite{Foulkes_01} 
which is somewhat less accurate but much cheaper. Systems with up to 
246 particles were simulated with the VMC algorithm, and the size 
dependence 
of the computed energies $E_N$
was modeled as
\begin{equation}
E_\infty = E_N + T_\infty - T_N +\beta/N, \label{eq:size_extrapolation}
\end{equation}
where $T_\infty$ and $T_N$ are the kinetic energy in the thermodynamic
limit and in the $N-$particle system (with TABC), respectively, and
$E_\infty$ and $\beta$ are fitting parameters. The $\chi^2$ value,
less than 2 on average, is at worst about 10 for 
$r_s\!=\!1$ and $\mu\!=\!4$, at $\zeta\!=\!0$. 
Figure~\ref{fig:size_extrapolation} shows the size extrapolation 
procedure for the Coulomb potential at $r_s\!=\!1$.
Since the dependence on spin polarization of the optimal value 
of $\beta$ is very weak (see Fig.~\ref{fig:size_extrapolation}),
a systematic study of the finite-size effect was carried out only 
for $\zeta\!=\!0$: for given $\mu$ and $r_s$, the same value of $\beta$, 
determined from the VMC energies of the paramagnetic fluid at several 
system sizes, was then used to estimate the finite-size correction to 
the DMC energy for both $\zeta\!=\!0$ and $\zeta\!=\!1$. For the unpolarized
gas, 
we found that the discrepancies on the correlation energy
with the coupled-cluster data of 
Refs.~\onlinecite{erf1} and~\onlinecite{erfgau}
are of the order of 5--8\%. 
\section{Analytic representation of the correlation energy}
\label{sec_fit}
\begin{figure}
\includegraphics[width=7.2cm]{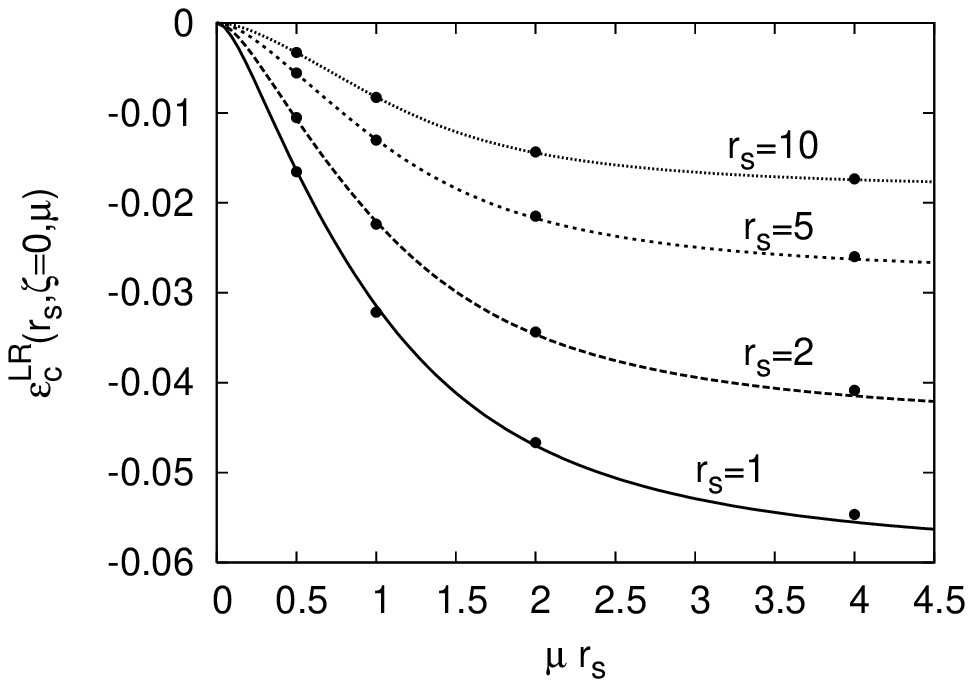} 
\includegraphics[width=7.2cm]{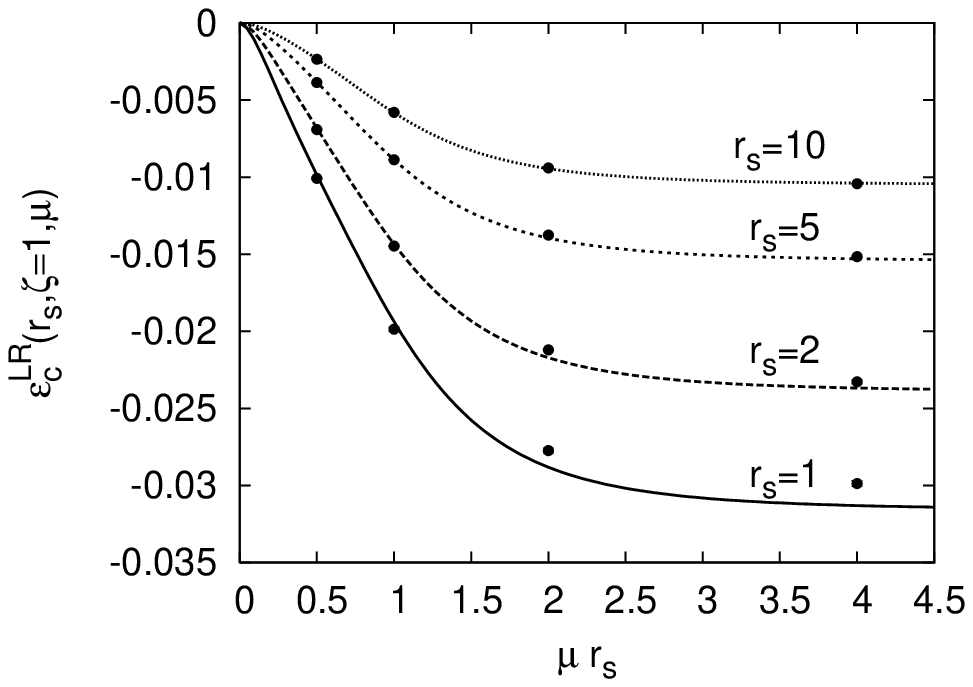} 
\caption{Our DMC data for the correlation energy ($\bullet$) of the electron gas
with long-range interaction $\erf(\mu r)/r$ are compared with the
fitting function (lines) of Eq.~(\ref{eq_fit}) for the unpolarized case (upper panel)
and the fully polarized case (lower panel). The statistical errors on
the DMC data are comparable with the symbol size.}
\label{fig_fit}
\end{figure}

We construct an analytical representation of the correlation energy as
\begin{eqnarray}
&&\eclr(r_s,\zeta,\mu) = 
\label{eq_fit}
\\
&& \frac{[\phi_2(\zeta)]^3Q\!\left(\displaystyle \frac{\mu\sqrt{r_s}}{\phi_2(\zeta)}\right)+a_1 \mu^3+a_2 \mu^4+
 a_3\mu^5+a_4\mu^6+a_5\mu^8}{(1+b_0^2\mu^2)^4},
\nonumber
\end{eqnarray}
where the function $Q$ is given by Eq.~(\ref{eq_Q}), 
the parameters
$a_i(r_s,\zeta)$ ensure the correct large-$\mu$ behavior of 
Eqs.~(\ref{eq_mugraz0})-(\ref{eq_mugraz1}), and
$b_0(r_s)$ is fixed by a best fit to our DMC data.
Some more free parameters which are adjusted to fit the DMC data are also contained in the coefficients $a_i(r_s,\zeta)$, whose definition requires a detailed explanation.
For the limits of Eqs.~(\ref{eq_mugraz0})-(\ref{eq_mugraz1}) we
use, for any spin-polarization $\zeta$, the approximation
\begin{eqnarray}
\eclr(r_s,\zeta,\mu)\Big|_{\mu\to\infty}  \!\!\!\!\!\!\approx && \ec(r_s,\zeta)
-\frac{3(1\!-\!\zeta^2)\,g_c(0,r_s,\zeta\!=\!0)}{8\,r_s^3\,\mu^2} \nonumber \\
 - && (1\!-\!\zeta^2)\frac{g(0,r_s,\zeta\!=\!0)}{\sqrt{2\pi}\, r_s^3\,\mu^3}
-\frac{9 \,c_4(r_s,\zeta)}{64 r_s^3\mu^4} \nonumber \\
 - && \frac{9\, c_5(r_s,\zeta)}{40\sqrt{2 \pi} r_s^3\mu^5} +
O(\mu^{-6})\, , \, {\rm with} \\
c_4(r_s,\zeta)  
= && \left(\frac{1\!+\!\zeta}{2}\right)^2g''\left(0,
r_s\left(\frac{2}{1\!+\!\zeta}\right)^{1/3}\!\!\!\!\!\!\!\!, \,\,\,\,\,\zeta\!=\!1\right)+ \nonumber \\
 + && \left(\frac{1\!-\!\zeta}{2}\right)^2g''\left(0,
r_s\left(\frac{2}{1\!-\!\zeta}\right)^{1/3}\!\!\!\!\!\!\!\!, \,\,\,\,\,\zeta\!=\!1\right) \nonumber \\
 + && (1\!-\!\zeta^2)D_2(r_s)-\frac{\phi_8(\zeta)}{5\,\alpha^2\,r_s^2}
\, \, , \, {\rm and} \label{eq_c4}\\
c_5(r_s,\zeta)  
= && \left(\frac{1\!+\!\zeta}{2}\right)^2g''\left(0,
r_s\left(\frac{2}{1\!+\!\zeta}\right)^{1/3}\!\!\!\!\!\!\!\!, \,\,\,\,\,\zeta\!=\!1\right)+ \nonumber \\
 + && \left(\frac{1\!-\!\zeta}{2}\right)^2g''\left(0,
r_s\left(\frac{2}{1\!-\!\zeta}\right)^{1/3}\!\!\!\!\!\!\!\!, \,\,\,\,\,\zeta\!=\!1\right) \nonumber \\
 + && (1\!-\!\zeta^2)D_3(r_s).
\end{eqnarray}
The function $\phi_8$ 
is defined by Eq.~(\ref{eq_phi}); $D_2(r_s)$ and $D_3(r_s)$ mimic the effect of the
$\ud$ correlation on the $\mu^{-4}$ and $\mu^{-5}$ large-$\mu$
coefficients, and are obtained by a best fit to the
DMC data. For the parallel-spin $g''(0,r_s,\zeta)$ and for the on-top
$g(0,r_s,\zeta)$ an exchange-like $\zeta$ dependence was assumed, starting
from the values at $\zeta\!=\!1$ and $\zeta\!=\!0$, respectively.
The on-top $g(0,r_s,\zeta\!=\!0)$ was taken from Ref.~\onlinecite{GP1},
while $g''(0,r_s,\zeta\!=\!1)$ was obtained as a best fit to our DMC data.
The parameters $a_i(r_s,\zeta)$ of Eq.~(\ref{eq_fit}) are then equal to
\begin{eqnarray}
a_1 & = & 4 \,b_0^6 \,C_3+b_0^8 \,C_5, \nonumber \\
a_2 & = & 4 \,b_0^6 \,C_2+b_0^8\, C_4+6\, b_0^4 \ec, \nonumber \\
a_3 & = & b_0^8 \,C_3, \nonumber \\
a_4 & = & b_0^8 \,C_2+4 \,b_0^6\, \ec \nonumber, \\
a_5 & = & b_0^8\,\ec, \nonumber
\end{eqnarray}
where $\ec(r_s,\zeta)$ is the
parametrization of the CA correlation energy 
as given by Perdew and Wang,\cite{PW92} and
\begin{eqnarray}
C_2 & = & -\frac{3(1\!-\!\zeta^2)\,g_c(0,r_s,\zeta\!=\!0)}{8\,r_s^3} \nonumber \\
C_3 & = &  - (1\!-\!\zeta^2)\frac{g(0,r_s,\zeta\!=\!0)}{\sqrt{2\pi}\, r_s^3} 
\nonumber \\
C_4 & = & -\frac{9\, c_4(r_s,\zeta)}{64 r_s^3} \nonumber \\
C_5 & =  & -\frac{9\, c_5(r_s,\zeta)}{40\sqrt{2 \pi} r_s^3}.
\label{eq_Ci} 
\end{eqnarray}
The functions $b_0$, $g''$, $D_2$
and $D_3$ are finally obtained from a best fit to the DMC data and read
\begin{eqnarray}
\phantom{\bigl[} b_0(r_s)  =   0.784949\,r_s  \\
\phantom{\Biggl[} g''(0,r_s,\zeta\!=\!1)  =   \frac{2^{5/3}}{5\,\alpha^2 \,r_s^2} \,
\frac{1-0.02267 r_s}{\left(1+0.4319 r_s+0.04 r_s^2\right)} 
\label{eq_gsec0} \\
\phantom{\Biggl[}D_2(r_s)  =  \frac{e^{-  0.547 r_s}}{r_s^2}\left(-0.388 r_s+0.676 r_s^2\right) \\
\phantom{\Biggl[}D_3(r_s)  =  \frac{e^{-0.31 r_s}}{r_s^3}\left(-4.95 r_s+ r_s^2\right).
\end{eqnarray}
Notice that, by our construction, Eq.~(\ref{eq_gsec0}) satisfies the exact high-density limit.\cite{rassolov}
Our data and the fitting function of Eq.~(\ref{eq_fit}) are shown in 
Fig.~\ref{fig_fit}. The small discrepancy at large $\mu$, 
particularly visible for $r_s=1$ and $\zeta=1$ on the scales of the figure,
is due to the condition that our fitting function recovers in the
Coulomb limit the Perdew-Wang parametrization\cite{PW92} of the CA
correlation energy, and it is consistent with our FN results being 
an upper bound to the data obtained\cite{CA} by CA using a nominally 
exact method.
\section{Pair-distribution functions and alternative separation of
exchange and correlation}
\label{sec_delta}
\begin{figure}
\includegraphics[width=7.2cm]{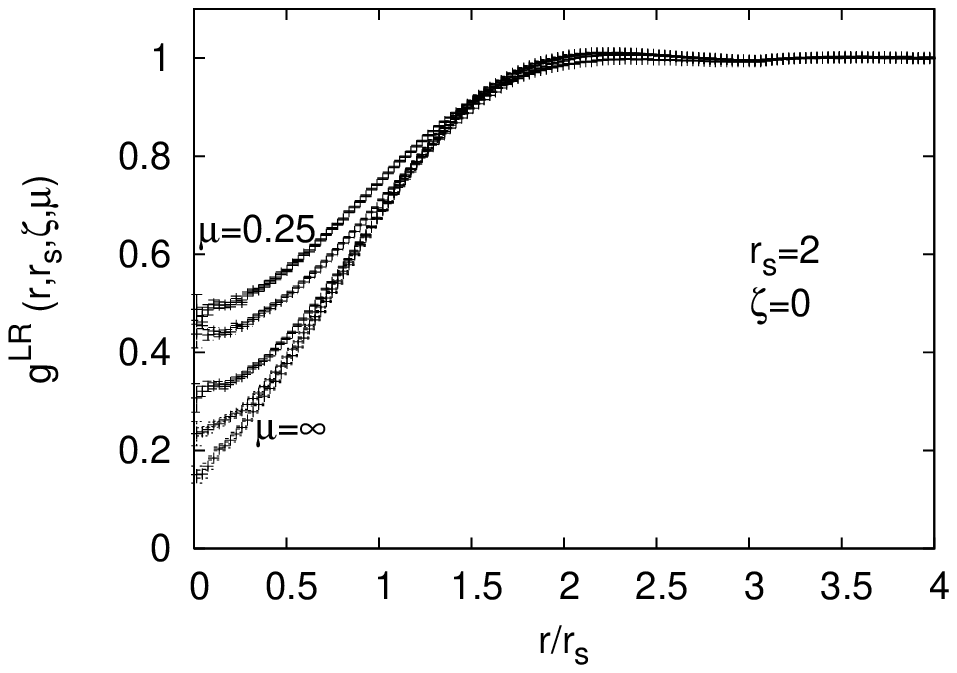} 
\includegraphics[width=7.2cm]{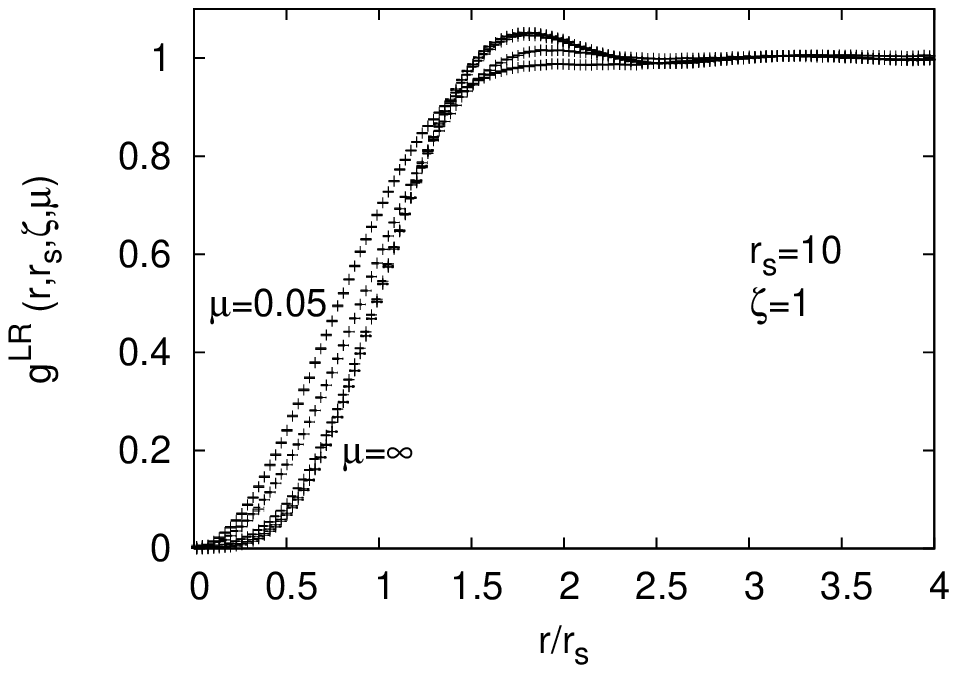} 
\caption{A sample of our DMC pair-distribution functions. Upper panel:
for $r_s\!=\!2$ and $\zeta\!=\!0$, $g^{\rm LR}$ is shown for $\mu\!=\!0.25, 0.5, 1, 2$, and for the Coulomb gas ($\mu\!=\!\infty$). Lower panel: for $r_s\!=\!10$ and $\zeta\!=\!1$,  $g^{\rm LR}$ is shown for $\mu\!=\!0.05, 0.1, 0.2, 0.4$, and for the Coulomb gas ($\mu\!=\!\infty$).}
\label{fig_gofr}
\end{figure}

From our DMC runs we also extracted,  in the usual way,\cite{OB} the 
pair-distribution functions $g^{\rm LR}(r,r_s,\zeta,\mu)$. 
A sample of our results is shown in Fig.~\ref{fig_gofr}. 
These functions are of interest in the framework  of the approach of 
Refs.~\onlinecite{erf1,conLDA,julGGA,sav_madrid,julien,janos,stoll}. While a local- (or local-spin-) density approximation for both exchange and correlation has been, and to a large extent still is, the most popular approach to Kohn-Sham calculations (possibly with GGA improvements), there is a growing interest\cite{ChiCitareEh} in optimized-effective-potential schemes,
where the exchange is treated exactly and the construction of approximations only concerns the correlation energy. The latter is naturally defined as whatever exceeds the exact-exchange energy, obtained from a single Slater determinant of Kohn-Sham orbitals. But once a multideterminantal, partially correlated wavefunction $\Psi^\mu$  [Eq.~(\ref{eq_LR-SR-spin})] is introduced, as in the modified schemes we are concerned with here, an alternative, more efficient choice, may be to construct approximations only for that portion of the correlation energy which is not already taken into account by $\Psi^\mu$. In other words, one may prefer to define\cite{TGS2} ``exchange'' and ``correlation'' energy functionals in the following way:
\begin{eqnarray}
\overline{E}_{x, {\rm md}}^\mu[n_\uparrow,n_\downarrow] & = &
\langle\Psi^\mu|V_{ee}-V_{\rm LR}^\mu|\Psi^\mu\rangle-
\overline{E}_{\rm H}^\mu[n] \label{eq_Exmd}\\
\overline{E}_{c, {\rm md}}^\mu[n_\uparrow,n_\downarrow] & = &
\overline{E}_{\rm xc}^\mu[n_\uparrow,n_\downarrow]-
\overline{E}_{x, {\rm md}}^\mu[n_\uparrow,n_\downarrow] 
\label{eq_Ecmd},
\end{eqnarray}
and then apply, e.g., the  LSD approximation only to the ``correlation'' energy functional of
Eq.~(\ref{eq_Ecmd}):\cite{TGS2}
\begin{equation}
\overline{E}_{c, {\rm md}}^\mu[n_\uparrow,n_\downarrow]=\int \!\! d\rv  \, n(\rv) \,\overline{\epsilon}_{\rm c,\;md}
(r_s(\rv),\zeta(\rv),\mu) . 
\end{equation}
Here
\begin{equation}
\overline{\epsilon}_{\rm c,\;md}(r_s,\zeta,\mu)= 
\ec(r_s,\zeta)-\eclr(r_s,\zeta,\mu)+\Delta_{\rm LR-SR}(r_s,\zeta,\mu),
\end{equation}
the mixed term $\Delta_{\rm LR-SR}(r_s,\zeta,\mu)$ is equal to
\beq
\Delta_{\rm LR-SR}(r_s,\zeta,\mu) =\!-\frac{n}{2}\int_0^\infty
\!\!\!\!\!\!\! 4\pi r^2 dr \,
g_c^{\rm LR}(r,r_s,\zeta,\mu)\, \frac{\erfc(\mu r)}{r},
\label{eq_deltadef}
\eeq
and $g_c^{\rm LR}$ is given by $g^{\rm LR}$ minus the
pair-distribution function of the noninteracting gas.
Using the results of Ref.~\onlinecite{GS3} it is easy to show
that, for large $\mu$, the mixed term $\Delta_{\rm LR-SR}$ behaves as
\begin{eqnarray}
\Delta_{\rm LR-SR}\Big|_{\mu\to\infty} \!\!\!\!\!\!\!\!\!\!\!\! & = & 
-\frac{3\,g_c(0,r_s,\zeta)}{8\,r_s^3\,\mu^2} \nonumber \\
& - & \frac{g(0,r_s,\zeta)(2\sqrt{2}-1)}{2\sqrt{\pi}\, r_s^3\,\mu^3}
+O(\mu^{-4})
\label{eq_deltamugraz0}
\end{eqnarray}
for $\zeta\neq 1$, and as
\begin{eqnarray}
\Delta_{\rm LR-SR}\Big|_{\mu\to\infty}  \!\!\!\!\!\!\!\!\!\!\!\! & = &
-\frac{9\,g_c''(0,r_s,\zeta\!=\!1)}{64\,r_s^3\,\mu^4} \nonumber \\
& - & \frac{3\, g''(0,r_s,\zeta\!=\!1) (3\!-\!\sqrt{2})}{20\sqrt{2\pi}\,r_s^3\,\mu^5}
+O(\mu^{-6})
\label{eq_deltamugraz1}
\end{eqnarray}
for $\zeta\!=\!1$, with the same notations of 
Eqs.(\ref{eq_mugraz0}-\ref{eq_mugraz1}).  \par In this Section we present an accurate parametrization of $\Delta_{\rm LR-SR}$.
Exploiting our DMC pair-distribution functions $g^{\rm LR}(r,r_s,\zeta,\mu)$,
we solved Eq.~(\ref{eq_deltadef}) by numerical integration, and
parametrized our results as
\beq
\Delta_{\rm LR-SR}=\frac{\delta_2\,\mu^2+\delta_3\,\mu^3+\delta_4\,\mu^4+
\delta_5\,\mu^5+\delta_6\,\mu^6}{(1+d_0^2\,\mu^2)^4},
\eeq
where the functions $\delta_i(r_s,\zeta)$ with $i=3...6$ guarantee
the correct large-$\mu$ behavior of Eqs.~(\ref{eq_deltamugraz0})-(\ref{eq_deltamugraz1}):
\begin{eqnarray}
\delta_3 & = & 4\,d_0^6\,\tilde{C}_3+d_0^8\,\tilde{C}_5\\
\delta_4 & = & 4\,d_0^6\,C_2+d_0^8\,C_4\\
\delta_5 & = & d_0^8\,\tilde{C}_3\\
\delta_6 & = & d_0^8\,C_2 \, .
\end{eqnarray}
Here $C_2(r_s,\zeta)$ and $C_4(r_s,\zeta)$ are those of
Eqs.~(\ref{eq_Ci});
\begin{eqnarray}
\tilde{C}_3 & = &  
-(1\!-\!\zeta^2)\frac{g(0,r_s,\zeta\!=\!0)(2\sqrt{2}-1)}{2\sqrt{\pi}\, r_s^3} ,
\nonumber \\
\tilde{C}_5  & = & -\frac{3\, c_5(r_s,\zeta)(3-\sqrt{2})}{20\sqrt{2 \pi} r_s^3};
\end{eqnarray} 
$g(0,r_s,\zeta\!=\!0)$ and $c_5(r_s,\zeta)$ are defined in
Sec.~\ref{sec_fit}. The remaining parameters $\delta_2(r_s)$ and
$d_0(r_s,\zeta)$ are fitted to our DMC data and read
\begin{eqnarray}
\delta_2(r_s) & = & 0.073867\,r_s^{3/2}\\
d_0(r_s,\zeta) & = & (0.70605+0.12927\,\zeta^2)\,r_s.
\end{eqnarray}
\section{Conclusions}
We have  presented a comprehensive numerical and analytic study of the ground-state energy of a homogeneous electron gas with
modified, long-range-only electron-electron interaction $\erf(\mu
r)/r$, as  a function of the cutoff parameter $\mu$, of the electronic 
density, and of spin polarization. The final outcome of this work is the publication of a reliable local-spin-density functional which fits the results of our quantum Monte Carlo simulations and automatically  incorporates exact limits. Such a functional (Sec.~\ref{sec_fit}), or its variant implying the use of an additional term also obtained in this work (Sec.~\ref{sec_delta}),
are the key ingredient for some recently proposed ``multideterminantal" versions of the density functional theory, where quantum chemistry and approximate exchange-correlation functionals are combined to optimally describe both long- and short-range electron correlations.\par
A fortran subroutine that evaluates our LSD exchange-correlation functional and
the corresponding potentials is available upon request to 
{\tt gori@lct.jussieu.fr}, or can be downloaded at 
{\tt http://www.lct.jussieu.fr/DFT/gori/elegas.html}.
\section*{Acknowledgments}
We thank A. Savin and J. Toulouse
for useful discussions, H. Stoll and J.G. \'Angy\'an for suggesting and
encouraging this work,
and gratefully acknowledge financial support from the Italian Ministry of Education, University and Research (MIUR) through COFIN 2005-2006 and the allocation of computer resources from INFM Iniziativa Calcolo Parallelo.

\appendix
\section{Details of Eq.~(\ref{eq_smallmu})}
\label{app_RPA}
We start from the RPA equations\cite{Ulf} for the spin-polarized electron gas,
and we simply replace the Coulomb interaction $1/r$ with the long-range
interaction $\erf(\mu r)/r$. We then repeat the analysis done in
Refs.~\onlinecite{WP91} and \onlinecite{WPHDsc} for the coulombic gas and find that, in the $r_s\to 0$ limit, the correlation energy is given by
\begin{eqnarray}
& & \eclr(r_s,\zeta,\mu)\Big|_{r_s\to 0}  =
\nonumber \\
& & -\frac{12}{\pi}
\int_0^\infty \frac{dy}{\alpha^2}\,y \int_0^\infty du\Biggl\{\alpha \, R_{\zeta}(u) \, 
e^{\textstyle\!-\!\left( \frac{y}{\alpha \mu \sqrt{r_s}}\right)^{\! 2}}
\nonumber \\
& &  - y^2\ln\left[1+\frac{\alpha}{y^2} \, R_\zeta(u) \,
e^{\textstyle\!-\!\left( \frac{y}{\alpha \mu \sqrt{r_s}}\right)^{\! 2\,\,}} \right]\Biggr\},
\label{eq_basicRPA}
\end{eqnarray}
where
\beq
R_{\zeta}(u)=\frac{1}{2}\left[z_1\, R\left(\frac{u}{z_1}
\right)+z_2\,R\left(\frac{u}{z_2}\right)\right],
\eeq
with
\beq
R(u)=\frac{1}{\pi}\left[1-u\arctan(u^{-1})\right],
\eeq
$z_1=(1\!+\!\zeta)^{1/3}$, $z_2=(1\!-\!\zeta)^{1/3}$, and
$\alpha=(4/9\pi)^{1/3}$. Equation~(\ref{eq_basicRPA})
already shows that the correlation energy becomes a function
of $\mu\sqrt{r_s}$ in the $r_s\to 0$ limit.\par
\begin{figure}
\includegraphics[width=8cm]{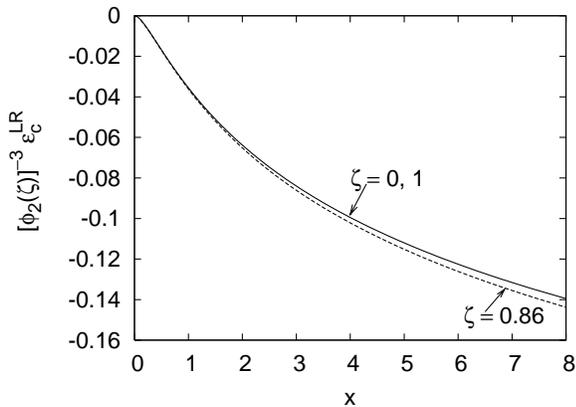} 
\caption{The numerical evaluation of Eq.~(\ref{eq_basicRPA}), as a function of
$x=\mu\sqrt{r_s}/\phi_2(\zeta)$, and multiplied
by $[\phi_2(\zeta)]^{-3}$. If the scaling of
Eq.~(\ref{eq_smallmu}) were exact, all the values corresponding
to different $\zeta$ would lie
on the solid curve. The value $\zeta=0.86$ shown in the figure corresponds
to the maximum deviation from the scaling of Eq.~(\ref{eq_basicRPA}).
}
\label{fig_rpa}
\end{figure}
To prove the small-$x$ behavior of Eq.~(\ref{eq_smallx}), take a value
of $\mu\sqrt{r_s}=a \ll 1$. In this case all the contribution
to the integral of Eq.~(\ref{eq_basicRPA}) comes from small $y$, since,
as soon as $y\gg a$, the integrand
goes to zero exponentially fast, as a function of $a$, when $a\to 0$ . We thus
integrate over $y$ Eq.~(\ref{eq_basicRPA}) between 0 and a value $q_1$ such
that $a\ll q_1\ll 1$. Since $y\ll 1$, the integral reduces to
\beq
-\frac{12}{\pi\alpha}\int_0^{q_1} dy\, y\, 
e^{-\left(\frac{y}{\alpha\,a}\right)^2} \int_0^\infty du\, R_{\zeta}(u),
\label{eq_provasmallx}
\eeq
which gives, to leading orders in $a$ when $a\to0$,
Eqs.~(\ref{eq_smallmu})-(\ref{eq_smallx}). The large $x$
behavior of Eq.~(\ref{eq_largex}) follows by considering the
$\mu\to\infty$ limit of Eq.~(\ref{eq_basicRPA}), which reduces to
the standard coulombic case studied in Ref.~\onlinecite{WPHDsc}.
\par
The $\zeta$ dependence of Eq.~(\ref{eq_smallmu}) is exact
in the $x\to 0$ limit of Eq.~(\ref{eq_provasmallx}). For larger $x$,
we evaluated Eq.~(\ref{eq_basicRPA}) numerically, and in Fig.~\ref{fig_rpa}
we report our results
multiplied by [$\phi_2(\zeta)]^{-3}$, as a function of 
$x=\mu\sqrt{r_s}/\phi_2(\zeta)$: if the scaling
of Eq.~(\ref{eq_smallmu}) were exact, all the values corresponding
to different $\zeta$
would lie on the solid curve, corresponding to $\zeta=0$ and 1. The
value $\zeta=0.86$ reported in the figure corresponds to the maximum
deviation from the scaling of  
Eq.~(\ref{eq_smallmu}), which is thus rather small. 
The function $Q(x)$ of Eq.~(\ref{eq_Q}) has been obtained
by fitting the RPA data of the solid curve. On the scale of Fig.~\ref{fig_rpa}
the fitting error is invisible.
\par
To conclude the discussion, we expect that the correlation energy
$\eclr$ lies on the curve of Fig.~\ref{fig_rpa} when $\mu r_s\ll 1$ 
(high-density or really long-range-only interaction on the scale $r_s$).
This means that at a given $r_s$, the ``exact'' $\eclr$ lies on
the curve of Fig.~\ref{fig_rpa} for values of $\mu$
such that $\mu\sqrt{r_s}\lesssim 1/\sqrt{r_s}$, that is, only the
densities $r_s\lesssim 0.1$ would be affected by the small deviations
from the scaling in $\zeta$ of Eq.~(\ref{eq_smallmu}), 
which appear at $x\gtrsim 3$.

\end{document}